\documentstyle[prb,aps,graphics]{revtex}

\newcommand{\FIG}[1]{}
\newcommand{\PREP}[1]{}

\newcommand{\vv}{\mbox{\bf v}}
\newcommand{\BB}{\mbox{\bf B}}

\begin{document}
\def\ltsima{$\; \buildrel < \over \sim \;$}
\def\gtsima{$\; \buildrel > \over \sim \;$}
\def\simlt{\lower.5ex\hbox{\ltsima}}
\def\simgt{\lower.5ex\hbox{\gtsima}}
\draft
\title{The two-dimensional magnetohydrodynamic Kelvin-Helmholtz instability: compressibility
        and large-scale coalescence effects}

\author{H. Baty}
\address{Observatoire Astronomique, 11 Rue de l'Universit\'e,
        67000 Strasbourg, France}

\author{R. Keppens}
\address{FOM-Institute for Plasma Physics Rijnhuizen, Association Euratom/FOM, 
        P.O.~Box 1207, 3430 BE Nieuwegein, The Netherlands} 

\author{P. Comte}
\address{Institut de m\'ecanique des fluides, 2 Rue Boussingault, \hfill \\
67000 Strasbourg, France}

\maketitle

\begin{abstract}
The Kelvin-Helmholtz (KH) instability occurring in a single shear flow
configuration that is embedded in a uniform flow-aligned magnetic field,
is revisited by means of high resolution two-dimensional (2D)
magnetohydrodynamic (MHD) simulations.
First, the calculations extend previous studies of magnetized shear
flows to a higher compressibility regime.
The nonlinear evolution of an isolated KH billow emerging from the
fastest growing linear mode for a convective sonic Mach number $M_{cs}=0.7$
layer is in many respects similar to its less compressible counterpart
(Mach $M_{cs}=0.5$). In particular, the disruptive regime where locally
amplified, initially weak magnetic fields, control the nonlinear saturation
process is found for Alfv\'en Mach numbers $4\simlt M_A \simlt 30$.
The most notable difference between $M_{cs}=0.7$ versus $M_{cs}=0.5$ layers
is that higher density contrasts and fast magnetosonic shocklet structures
are observed.
Second, the use of adaptive mesh refinement allows to
parametrically explore much larger computational domains, including up to
$22$ wavelengths of the linearly dominant mode. A strong process of large-scale
coalescence is found, whatever the magnetic field regime. It proceeds
through continuous pairing/merging events between adjacent
vortices up to the point where the final large-scale vortical structure
reaches the domain dimensions. This pairing/merging process is attributed
to the growth of subharmonic modes and is mainly controlled by relative phase
differences between them. 
These grid-adaptive simulations demonstrate that even in very weak magnetic
field regimes ($M_A \simeq 30$), 
the large-scale KH coalescence process can trigger tearing-type
reconnection events previously identified in cospatial current-vortex sheets.
\end{abstract}


\section{Introduction}\label{s-intro}

The Kelvin-Helmholtz (KH) instability in sheared flow configurations
is an efficient mechanism to initiate mixing of fluids,
transport of momentum and energy, and the development of turbulence.
In many astrophysical or geophysical environments, magnetized
shear flow layers with transonic or supersonic velocities
are of concern. For example, this is the case for
astrophysical jets, where
supersonic magnetized flows emanate from young stellar
objects or active galactic nuclei.
A full understanding of the various nonlinear processes playing a role
in the development of the KH instability in magnetohydrodynamics (MHD)
is thus of prime importance.

In its most basic configuration, an MHD study considers a shear layer
separating two conducting fluids flowing at different speeds with an initially
homogeneous magnetic field.
The linear stability of such a uniformly magnetized shear layer
is well known from the pioneering studies in incompressible MHD
by Chandrasekhar.~\cite{cha61} Miura and Pritchett~\cite{miu82} have extended
the linear stability study to the compressible case.
Stability crucially depends on both the strength and orientation
of the magnetic field vector with respect to the velocity. Depending on this
relative orientation, two classes of configurations are generally considered:
a parallel and a transverse one.

In transverse configurations, the nonlinear evolution of the KH instability
has been simulated numerically, for different
geophysical/magnetospheric applications (see Miura~\cite{miu97} 
and references therein).
In the two-dimensional (2D) transverse case,
if the wavenumber parallel to the magnetic field vanishes,
only slight differences are obtained compared to a pure hydrodynamic
configuration. This is because the effect of the magnetic field then
appears only through an additional magnetic pressure
term in the total pressure, while the line bending term vanishes.
Hence, only the fast magnetosonic mode is excited
in MHD, whereas it is the sonic one in hydrodynamics.

A parallel configuration allows a much richer complexity in the nonlinear
evolution of the MHD KH instability. This has been investigated for a
shear layer embedded in a uniform magnetic field, both in 2D,
~\cite{mal96,fra96,jon97,kep99}
and more recently in 2.5D and 3D simulations.~\cite{jeo00,ryu00,ket99}
The nonlinear behaviour varies dramatically with the Alfv\'en Mach number
$M_A \equiv V/v_A$ of the background configuration, where $V$ is the total
velocity jump across the shear layer and $v_A$ is the Alfv\'en speed.
Three magnetic regimes, briefly discussed in section~III, that characterize the
development of KH instabilities have been identified.
However, most of the studies mentioned above were restricted to
subsonic/transonic layers where the sonic Mach number of the velocity
transition across the shear layer is
$M_s\equiv V/c_s \le 1$ (for sound speed $c_s$). 
Moreover, to avoid excessive computational costs, earlier work
typically considered a periodic section of the flow
having a length equal to the wavelength of the linearly fastest growing mode.
While this allowed for detailed modeling of nonlinear, magnetically
controlled breakup of a single
KH billow, any trend to large-scale coalescence is excluded from these models.
Such coalescence due to pairing/merging of
adjacent vortices has been reported in 2D transverse configurations
considering longer periodic sections,~\cite{miu97,miu99}
as well as in purely hydrodynamic
simulations.~\cite{les88,les97}
While these longitudinally periodic studies focus on the temporal development,
coalescence has also been found in spatially developing
instabilities, convected with the flow.
This was obtained in a magnetospheric context by Wu~\cite{wu86}
and Manual and Samson,~\cite{man93} who
reported the formation of large vortices due to large-scale coalescence
effects in 2D non periodic simulations.

The aim of the present paper is therefore twofold: (1) to extend single billow
studies of a 2D uniform parallel configuration to higher flow velocities where $M_s \ge 1$;
and (2) to investigate the effects of large-scale coalescence by using very long
periodic sections of the flow. The latter significantly benefits from
the use of adaptive mesh refinement in our computations,
since it allows us to simulate large spatial domains
without prohibitive computational
costs.

The paper is
organized as follows. The physical model and the numerical codes are presented
in section~II. In section~III, we show the results of the single vortex
evolution in supersonic versus transonic layers.
Section~IV focuses on the large-scale coalescence process
that occurs in longitudinally extended shear layers.
Finally, we conclude and briefly discuss the consequences of our
findings in the context of astrophysical jets.

\section{The physical problem}

\subsection{The MHD equations and initial configuration}

The linear and early nonlinear phases of the KH instability are
ideal for the parallel configuration studied here.  Hence, we consider
plasmas with very large kinetic and magnetic Reynolds numbers,
as is indeed the case for many astrophysical environments.
However, non zero viscosity and
resistivity is important in later evolutionary stages
allowing for momentum exchange and magnetic reconnection, respectively.
In most of this work,
the set of ideal MHD equations is solved numerically, relying
on the inherent numerical resistivity and viscosity to mimick
dissipative processes (see the discussion in Jones {\it et al.}~\cite{jon97}).
Selected cases will make use of fully resistive MHD.
The ideal compressible MHD equations can be written in conservative form as
\begin{equation}
\frac{\partial \rho}{\partial t}+\nabla \cdot (\rho \vv)=0,
\end{equation}
\begin{equation}
\frac{\partial (\rho \vv)}{\partial t}+ \nabla \cdot [ \rho \vv \vv + p_{tot} I-
\BB \BB]=0,
\end{equation}
\begin{equation}
\frac{\partial e}{\partial t}+ \nabla \cdot (e \vv) + \nabla \cdot (p_{tot} \vv) - \nabla \cdot (\vv \cdot \BB \BB )= 0,
\label{q-e}
\end{equation}
\begin{equation}
\frac{\partial \BB}{\partial t}+
\nabla \cdot (\vv \BB-\BB \vv)= 0.
\label{q-b}
\end{equation}
Here, $\rho$ is the mass density, $\bf{v}$ the fluid velocity, $\bf{B}$
the magnetic field, and $t$ time. $I$ is the identity tensor.
We have defined the total plasma pressure $p_{tot} = p + \frac{1}{2} B^2$,
where the thermal pressure $p$ is related to the energy density $e$
as $p = (\gamma - 1)(e - \frac{1}{2}\rho v^2 - \frac{1}{2}B^2)$.
We set the ratio of specific heats $\gamma$ equal to $5/3$. Our units
are such that the magnetic permeability is one.

We solve the above set of nonlinear equations as an initial value problem
in two spatial dimensions and cartesian geometry $(x,y)$.
In the initial background configuration, the fluid moves along the longitudinal
$x$ direction with a velocity $v_x$ given by
\begin{equation}
v_x(y) = \frac{V}{2} \tanh \left ({\frac{y} {a}} \right ),
\end{equation}
where $a$ is the half-width of the shear layer situated at $y = 0$.
The $y$ direction is the cross-stream, transverse direction.
Note that the interface is at rest: this choice of reference
frame is well adapted to the study of the KH development, as the
instability is advected at the local flow velocity (a statement exactly true
for a subsonic flow layer). The initial magnetic field is aligned
with the flow and has a uniform strength given by $B_x = B_0$.
The initial thermal pressure $p_0$ and density $\rho_0$ are set equal to one,
defining our normalisation. Consequently, the sonic speed is uniform
and $c_s = (\gamma p /\rho)^{1/2} = 1.29$. In the present study, we fix
$a = 0.05$ and the dimensions of the computational domain $L_x \times 2L_y$
can vary from case to case. In our units, the growth times of the linearly
fastest modes are thus typically $t_g \simeq 1.7$ (taking $t_g \equiv \Gamma^{-1}$
where $\Gamma$ is the linear growth rate).~\cite{kep99}

\subsection{The numerical procedure}

We calculate the evolution of the KH unstable layers with
the general finite-volume based Versatile Advection Code (VAC, see
{\tt http://www.phys.uu.nl/}$\sim${\tt toth})~\cite{tot96} and its
recent grid-adaptive variant AMRVAC.~\cite{kep03}
The latter uses an automated Adaptive Mesh Refinement (AMR) strategy,
where a base grid is refined by adding finer level grids
where a higher resolution is needed. Finer level grids are adjusted, inserted
or removed by periodically checking if the grid structure should be altered
in response to the flow dynamics.
This procedure allows us to follow shock-dominated or
coexisting global and local plasma dynamics accurately in a much
more efficient way than with a global refinement of a static grid.

All simulations make use of a second-order time accurate
shock-capturing method employing
a Roe-type approximate Riemann solver, namely an explicit
one-step total variation diminishing (TVD) scheme with minmod limiting
on the characteristic waves.~\cite{col84,har83}
In AMRVAC, the automated AMR strategy allows
for a grid-level dependent choice of the spatial discretization. We follow
Keppens {\it et al.},~\cite{kep03} and use the computationally beneficial combination of
a robust two-step Total Variation Diminishing Lax-Friedrichs method on
all but the finest grid level, together with the less diffusive
TVD scheme on the highest one. Note that all discretization methods
in VAC, and all combinations of grid level dependent spatial discretization
methods in AMRVAC, are fully conservative. This implies that when the
boundary conditions are conservative as well, exact conservation of mass,
momenta, total energy, and magnetic flux holds for ideal MHD simulations.
To handle the solenoidal constraint on the magnetic field
$\nabla \cdot \BB=0$, our VAC simulations apply a projection scheme at
every time step in order to remove any numerically generated
divergence of the magnetic field up to a predefined accuracy.~\cite{bra80}
In grid-adaptive simulations, this is handled by a diffusive source term
treatment which damps the errors at their maximal rate in accord with
the prevailing Courant-Friedrichs-Lewy condition. This was shown to be
effective for multi-D AMR MHD simulations.~\cite{kep03}
AMRVAC results employed four grid levels
with a refinement factor of $2$ between two consecutive levels.
Refining was done based on a Richardson-type extrapolation procedure,
using a weighted average of density, $x$-momentum, and longitudinal $B_x$
components.

We assume periodicity along the longitudinal $x$ direction. The consequence is
to restrict the longitudinal wavelengths $\lambda$ to those
Fourier components $\lambda = L_x/n$, $n$ being an positive integer,
that can fit in the box length $L_x$. Taking $L_x$ equal to the
wavelength of the linearly
fastest growing mode $\lambda_m$
prohibits the formation of structures on a larger scale.
Large-scale coalescence is allowed by taking $L_x$
greatly exceeding $\lambda_m$.
We use free outflow boundaries on the lateral sides at $y = \pm L_y$.

The initial configuration is perturbed with two
different forms for the $t=0$ transverse velocity.
For deterministic runs, we follow Keppens {\it et al.}~\cite{kep99} and
use the functional form
\begin{equation}
v_y = \delta V \exp \lbrack -\left(\frac{y}{4a}\right)^{2} \rbrack
\times \sin\left(kx\right),
\label{deter}
\end{equation}
where $k = 2 \pi / \lambda$ is the longitudinal wavenumber. We take a small
amplitude $\delta V = 0.01$.
This expression corresponds to a normal mode form with a Gaussian
decay in the transverse direction. A second perturbation used
in this work is
\begin{equation}
v_y = \delta V \exp \lbrack -\left(\frac{y}{4a}\right)^{2} \rbrack
\times ran(x),
\label{random}
\end{equation}
where the function $ran(x)$ represents
a random number generator in the range [-0.5:0.5].
This form corresponds to white noise and is appropriate to
let all unstable wavelengths grow in a natural way, in configurations having
a long computational domain length that greatly exceeds $\lambda_m$.

\section{Numerical results: single vortex}

\subsection{Previous studies: subsonic/transonic layers}

We briefly recall previous results on the
single vortex formation associated with the linearly fastest growing
KH mode in 2D subsonic/transonic layers 
where $M_s \le 1$.~\cite{mal96,fra96,jon97,kep99}
The flow is {\em linearly stable} if the magnetic tension of
the initial background magnetic field is so strong that it prevents the
development of a small perturbation of the shear layer.
This linear stability holds for $M_{A} \simlt 2$.
When $4 \simgt M_{A} \simgt 2$, the initial field can not
provide linear stability, but the flow can be nonlinearly stabilized by the
enhanced magnetic tension generated by the perturbation.
This first regime is referred to as the {\em strong field} or the {\em
nonlinearly stable} regime, as the instability is stopped
after a small amount of quasi-linear growth.
In a weaker magnetic field configuration
the shear layer can roll up leading to the formation of a vortex aligned
with the flow, similar to an unmagnetized case.
Meanwhile, the magnetic field lines are expelled from the
vortex center, stretched, and amplified around the vortex perimeter.
A nonlinear saturation then occurs when the magnetic field becomes
locally dominant, i.e. when the field line tension is able to overcome
the centrifugal force associated to the vortical motion.
At this point, a structure similar to the classical
Kelvin's cat's eye (of pure hydrodynamics) is formed, enriched by local
strands of strongly amplified magnetic fields.
This also leads to magnetic reversals, so that the cat's eye structure
is subsequently rapidly disrupted due to magnetic reconnection in fully
resistive MHD. This {\em weak field} regime is qualified as
{\em disruptive}, and occurs for $20 \simgt M_A \simgt 4$.
Finally, a third regime is obtained when $M_A \simgt 20$.~\cite{jon97}
The development of KH instabilities in this {\em very
weak field} regime is essentially hydrodynamic in the early stages, and a nearly
classical hydrodynamic cat's eye is formed. Later, the role of
the magnetic field is to enhance its slow dissipation.

The first two magnetic regimes end up in a relaxed state with
an enlarged (in the cross-stream direction)
central flow layer of heated and lower density plasma.
In this quasi-steady laminar endstate, the velocity and magnetic field vectors
are nearly aligned everywhere.
The longitudinal kinetic energy, which feeds the instability, is thereby reduced
as compared to its initial level. In the strong field case, this is
achieved through local amplification and stretching of the magnetic field,
with a weak exchange between the different forms of energy.
In cases where the vortex is disrupted by reconnection (weak field regime),
the total (kinetic+magnetic) energy stored by the vortical motion is fully
released on a rapid time scale. For the very weak field regime, the vortex
can survive and decays at a slow rate through viscous and resistive
dissipation effects.

\subsection{Extension to higher compressibility regimes}

We extend the results obtained for the subsonic/transonic
regime ($M_s \le 1$) to a `supersonic' flow layer with $M_s\ge 1$. 
All our nonlinear
simulations reported below refer to $M_s=1$ layers as `transonic' and will
go up to slightly
`supersonic' layers at Mach number $M_s = 1.4$.
As the flow velocity increases, the linear stability of the shear layer
is enhanced. Typically, the growth rate is reduced by a factor of order
$10$ for $M_s = 2$ when compared to sonic Mach numbers smaller than one,
for Alfv\'en Mach numbers $M_A \gg M_s$.~\cite{kep99}
This stability enhancement is due to an increased intrinsic 
compressibility, usually
characterized using the sonic convective Mach number $M_{cs}$ (Mach number in the frame convected at the phase velocity of the unstable KH wave).~\cite{miu92}
In our configurations, the `transonic' $M_s = 1$ 
and `supersonic' $M_s = 1.4$ layers
correspond to subsonic $M_{cs} = M_s/2 = 0.5$ and $0.7$ flows, respectively,
and mainly differ in their intrinsic compressibility.
In fact, the 2D `classical' KH mode is stabilized at a critical $M_s^{\rm cr}$
value close to $2.5$ when $M_A = 10$ (see Fig.~2 in Miura~\cite{miu90}).
For a pure hydrodynamic shear layer, this stability criterion is $M_s \ge
2 \sqrt{2}$.~\cite{blu70} While full stability is
obtained for the vortex sheet (i.e. in the limit of vanishing layer
width $a$) when $M_s>M_s^{\rm cr}$,
new supersonic oscillatory instabilities appear in
a finite-width layer above this threshold.~\cite{blu75,dra77,miu90,fer83,ray82}
Selected runs at $M_s=3$ confirmed the existence of these unstable traveling
modes, but the perturbations attain a rather low level in the non linear phase,
consistent with earlier findings by Miura.~\cite{miu90} Hence, in very 
supersonic layers, the dominant instability
is likely to be less dangerous for the integrity of the background flow.
Consequently, we will not consider this truly supersonic regime (where
also $M_{cs}\ge 1$) further in this work.

In this section, a domain length $L_x=1$ is chosen in order to
allow the growth of a single periodic structure at a wavelength
$L_x \approx \lambda_m$.
Strictly speaking, the wavelength of the linearly fastest
growing mode $\lambda_m$ varies
with the fast magnetosonic Mach number $M_f = V/(c_s + v_A)^{1/2}$.
However, this dependence
is weak for the range of Mach numbers considered here (see for example Fig.~3
of Keppens {\it et al.}~\cite{kep99}, and Figs.~4 and 5 of 
Miura and Pritchett~\cite{miu82}).
Note that this domain choice $L_x=1$ imposes a minimum
wavenumber value included in the simulation at $ka = 0.314$.
We set $L_y = 1$ and perform VAC simulations at
a resolution of $200\times 400$, previously
shown to be sufficient to follow the nonlinear
saturation phases.~\cite{kep99} For a few $M_A$ values,
we repeated the runs with up to $350\times700$ grid cells
in order to check the convergence of our results.
The perturbation used to initiate the
instability is the deterministic one given by Eq.~(\ref{deter}).

First, we make a global comparison beween the development of the KH instability
in supersonic $M_s=1.4$ versus transonic $M_s=1$ layers.
In Fig.~\ref{fig1}, one can see the typical time evolution
of the transverse kinetic
$e_{k,y}$, and magnetic $e_{m,y}$ energies, as well as the evolution of
the longitudinal kinetic energy $E_{k,x}$ for three radically different magnetic
field strengths. These energies are volume averaged quantities defined as
\begin{equation}
e_{k,y} =
\frac{1}{V_b} \int\limits_{V_b} \frac{\rho V_y^{2}}{2} \,dx dy,
\end{equation}
\begin{equation}
e_{m,y} =
\frac{1}{V_b} \int\limits_{V_b} \frac{B_y^{2}}{2} \,dx dy,
\end{equation}
\begin{equation}
E_{k,x} =
\frac{1}{V_b} \int\limits_{V_b} \frac{\rho V_x^{2}}{2} \,dx dy,
\end{equation}
where $V_b$ is the box volume, namely $V_b = 2 L_y \times L_x$. The
transverse energies $e_{k,y}$ and $e_{m,y}$
serve as a measure of the energy built up by the instability through
the vortical motions, while the
longitudinal kinetic energy in the initial background flow
feeds the instability.

For a small Alfv\'en Mach number $M_A = 3.33$, in the strong field
regime, the kinetic and magnetic
energies attain a maximum value at roughly the same time, before
decreasing monotically towards zero. At higher Alfv\'en Mach number
values ($M_A = 10, 100$), a maximum in kinetic energy is obtained first,
and after a small time delay is followed by a maximum in magnetic energy.
For $M_A = 10$, these energies are released after the occurrence of several
secondary peaks that correspond to the formation of smaller secondary vortices.
Generally, the first peak in $e_{k,y}$ corresponds
to the non linear saturation of the vortex, while the first maximum
in transverse magnetic energy $e_{m,y}$
indicates the beginning of the disruption process due to magnetic reconnection.
\cite{kep99} The smaller vortices formed later are also disrupted by
secondary reconnection events.~\cite{fra96}
For the very weak field case having $M_A = 100$,
the energies slowly decay in an oscillatory way
after the saturation phase, as the cat's eye structure persists
without being disrupted.
The period of oscillation is simply the rotation time of the vortical flow.
As seen in Fig.~\ref{fig1},
these observations hold for both transonic and supersonic cases, so that
slightly supersonic and transonic 
layers appear to behave very similar.
The same conclusion can be reached
by comparing the time evolution of other characteristic quantities.

The three regimes (very weak field, disruptive, and strong field) identified for
transonic shear layers were initially deduced from
simple but highlighting estimates, based on a few runs only.~\cite{jon97}
In order to precisely quantify
the transition between the three regimes, we performed a parametric
study for transonic as well as supersonic shear layers.
We ran a very large number of simulations with
Alfv\'en Mach numbers taken in the range $[2:160]$.
All these results are reported in Fig.~\ref{fig2}, which shows
the dependence of the level of the previously defined energy peak
values on the magnetic field strength.
This figure confirms that the ranges of Alfv\'en Mach numbers defining
the three magnetic regimes are similar for supersonic and transonic layers.
In fact, when normalized by the background kinetic energy,
the perturbed saturated energy levels as well as the total amount of
longitudinal kinetic energy released are reduced by a few percent only in
supersonic versus transonic cases.
The transitions between the successive regimes are not sharp.
From Fig.~\ref{fig2} one can deduce that the transition between
the disruptive and dissipative regime occurs at an Alfv\'en Mach number
that is closer to $M_A = 30$ than to $M_A = 20$.
Furthermore, on close inspection of the disruptive regime in Fig.~\ref{fig2},
a pronounced maximum in the perturbed magnetic energy clearly appears for
$M_A = 8$. The cat's eye structure with spiral arms obtained
from density (see Fig.~\ref{density}) and magnetic maps indicate that
the initial flow interface is rolled-up
by $1/2$ turn for $M_A = 5$, by $1.5$ turn for $M_A = 15$, and by exactly
one turn for $M_A = 8$. At the transition
Mach number $M_A = 30$, this rolling-up value is close to $2$ turns.
Note that for field strengths higher than $M_A = 8$, the levels
reached in transverse kinetic and magnetic energies at saturation are
comparable, in contrast to weaker field cases. The maximum
magnetic field measured at saturation for the disruptive regime indicates that the amplification
factor is closer to $M_A$ than to $M_A/2$ (obtained by Miura~\cite{miu84}
in pioneering MHD simulations).

Figure~\ref{density} shows the density distribution at the time of saturation
for 4 selected cases. We contrast a disruptive case $M_A=15$ with a very
weak field case $M_A=100$ for transonic versus supersonic $M_s=1.4$ layer.
Notable differences between supersonic and transonic simulations are
the appearance of shocks in the supersonic case.
Furthermore, the vortex is a little flatenned for slightly supersonic
layer $M_s = 1.4$, an effect known to arise in pure hydrodynamics
due to compressibility for $M_{cs} \simgt 0.5-0.6$.~\cite{lel89}
Compressive effects are also responsible for a pronounced difference in
the density contrast during the entire evolution:
it is higher by roughly $50$ per cent for $M_s = 1.4$ versus $M_s = 1$.

The shocks that develop in supersonic layers form at the periphery
of the vortex, remain attached to it, and are nearly aligned with the
transverse direction. They slowly
travel against the background flow, and
disappear soon after the saturation stage for the supersonic $M_A=15$ case.
For more strongly magnetized supersonic layers ($M_A \simlt 13$) no
such transient shocklets form.
In the very weak field
nearly hydrodynamic case $M_A=100$, the shocks persist much longer.
In the disruptive $M_A=15$ case, the shock speeds were found
to be $s \simeq \pm 0.2$ at the time of saturation $t=4.26$ shown in
Fig.~\ref{density}.
This corresponds to plasma velocities that cross the fast magnetosonic
$M_f=1$ transition in the co-moving frame, confirming that
these shocks are of fast magnetosonic type. Similar eddy shocklets are well
known to arise in pure hydrodynamics.~\cite{lel89} 
We show in 
Fig.~\ref{figshock} several thermodynamic and
magnetic quantities along a 1D cross section in the $x$-direction for
$M_A=15$ (cross section at $y=0.3$ and $t=4.26$ as in Fig.~\ref{density}),
and similarly for $M_A=100$ (at $y=0.2$ and $t=4.26$, slightly later than
the $t=4$ frame shown in Fig.~\ref{density}). Both shocks are very weak,
since the entropy increase is of order 1 pro mille for $M_A=100$, and one order
of magnitude smaller for $M_A=15$. The shock speed can be determined from
the Rankine Hugoniot relation across the shock, and is printed in the figure.
The magnetic field components change little across the shock (note the
multiplicative factors used to show their slight variations), 
but their variations are consistent with a minute bending of the field away 
from the shock normal
(as expected for a fast magnetosonic shock). One must take into account the
fact that the shock front is slightly bent backwards for $M_A=15$ and
forwards for $M_A=100$ at that position and time. The flow remains 
superAlfv\'enic in the co-moving frame. As a final note, the shock structure
for the more strongly magnetized cases shows evidence of a more 
complex structure of several shock segments where the shock meets the
low density vortex perimeter.

\section{Numerical results: N vortices}

\subsection{Many vortices}

The motivation to study multiple vortices arises from the known trend to
large-scale coalescence in the 2D transverse case.~\cite{miu97,miu99}
Until now, claims in the literature exist that no coalescence is found
for the parallel configuration.~\cite{fra96}

Our results obtained for a large simulation domain take a
rectangular box of size of $L_x \times 2L_y = 10\times8$.
The initialisation for the perturbations
makes use of the random noise form given by Eq.~(\ref{random}).
AMRVAC is used with a base resolution of $100\times100$ grid cells and three
finer levels of refinement, effectively achieving a resolution of
$800\times800$.
We again investigate two $M_s$ values equal to $1$ and $1.4$.
For each sonic Mach number, up to $5$ Alfv\'en Mach
numbers equal to $M_A = 3.33, 7, 15, 30,$ and $100$ are considered.

The results for the time evolution of the longitudinal $E_{k,x}$
and transverse kinetic $e_{k,y}$ energies are plotted in Fig.~\ref{tergnvor} for $3$
magnetic field strengths in the transonic layer. 
Plotting the same indicators for the supersonic runs yields
very similar trends. The deviation from exact total energy
conservation in these long time runs with AMRVAC is only a few percent,
and since AMRVAC is fully conservative,
entirely due to the treatment of the lateral $y=\pm L_y$ boundaries
as open.

Let us, first, discuss the results obtained for the very weak field regime with
$M_A = 100$. As seen in Fig.~\ref{tergnvor}, the characteristic time evolution
of the energetic quantities is fundamentally different
from that of the single vortex.
Indeed, after an initial phase of exponential growth of $e_{k,y}$ that ends at
$t\simeq 7$, a phase of continuous global increase is obtained.
The transverse energy rises approximately linearly with a superposed
oscillation. Corresponding snapshots
displaying the density structure
at different times are reported in Fig.~\ref{coalm1ma100}.
At early times close to the first local maximum of $e_{k,y}$, we can identify
$11$ successive vortices.
This number of vortices is roughly in accord with the wavelength of the most
unstable mode, which grows naturally from the added noise.
For the same reason, the vortex strength can differ from one vortex to the
next. The second snapshot in Fig.~\ref{coalm1ma100} corresponds
to a time close to the second $e_{k,y}$ maximum, and one can
easily see that two pairing/merging events have already occured
leading to $9$ remaining vortices. This indicates the start of
a continuous process of successive pairing/merging events, which ends
up only when a big single vortex is formed at $t \simeq 45$.
For the simulation shown in Fig.~\ref{coalm1ma100}, the number of
vortices changes as follows:
$11 \rightarrow 9 \rightarrow 7 \rightarrow 5 \rightarrow 4 \rightarrow 3
\rightarrow 2 \rightarrow 1$. Each merger corresponds to a
successive local maximum of $e_{k,y}$ in Fig.~\ref{tergnvor}.
During the same time, the longitudinal
kinetic energy is continuously decreasing in the same oscillatory way.
At the end of our simulation, the characteristic cross-stream scale length
of the final vortex,
as well as the attained level in transverse kinetic energy, is
roughly $10$ times larger than those deduced for the initial small vortices.
This final big single vortex subsequently decays at a low rate through
viscous/resistive dissipation.

A similar qualitative evolution is obtained for the supersonic very weak
field counterpart case, illustrated in Fig.~\ref{coalm14ma100}. However,
the merging is leading to complex eddy shocklet interactions
which were previously observed in the single vortex evolution. As a consequence,
higher density contrasts can be reached. In the final snapshot shown, the
larger structures show clear evidence of incomplete merging events due to
the various shock interactions. Hydrodynamic simulations also demonstrated
similar shock-shock interactions.~\cite{fu97}

For the strong field regime with $M_A = 3.33$, the time
evolution of the energy monitors (Fig.~\ref{tergnvor})
is again very different from the single vortex one (Fig.~\ref{fig1}).
Indeed, after a saturation phase at $t \simeq 9$,
the transverse kinetic energy shows a long phase of decrease till
times $t\approx 20$.
The snapshots showing the density structure in Fig.~\ref{coalm1ma3},
indicate that the shear layer is initially deformed with a growing perturbation
having $11$ wavelengths. The perturbation strength is irregular
along the layer. Subsequently, a saturation occurs with at the same time
$2$ merging events between $2$ adjacent wavelength structures, at
$x\approx 2$ and $x\approx 6.5$. From times $t=10$ till $t=15$, the
evolution is similar to that of a single-vortex simulation with $L_x = 1$, with
a relaxation phase during which an enlarged shear layer tends to form.
Suddenly, at time $t\simeq 20$, the density shows evidence of a larger scale
structure along the flow shear layer. This leads to
exponential rebirth of a sinusoidal-like perturbation having
two longitudinal wavelengths of different amplitude. A final merging event
leads to a single wavelength structure remaining at $t \simeq 50$.
A subsequent relaxation then occurs, as seen in the decreasing transverse
energy (Fig.~\ref{tergnvor}).
Our simulations show a similar scenario for the supersonic layer
counterpart of this magnetic regime.
Again, the characteristic lengthscales of the final
single perturbation are about ten times those of a small single one.

Let us now turn to simulations for
an intermediate magnetic field strength with $M_A = 7$, in
the disruptive regime of the single MHD vortex.
For early times $t \simlt 9$, in a way similar to
the previous case, two pairing/merging events can be identified to occur
once $11$ or $12$ initial vortices have developed.
As seen in Fig.~\ref{coalm1ma7}, the
subsequent evolution is characterized by a saturation phase
in which the remaining vortices
become elongated. This is followed by the beginning of a disruption process,
clearly visible in the snapshot taken at $t= 9.1$.
Whilst this magnetic reconnection proceeds locally,
a few merging/pairing events are also taking place. The reconnection is unable
to fully destroy individual vortices, which continue their trend to
pair and merge.
The distorted layer shows the rebirth of $4$ vortices
of unequal strength (see the snapshot taken at $t=13.1$). Later, a pairing
between two vortices occurs near $x=8$ whilst the two other vortices are
partially disrupted by reconnection. Finally, the cascade towards
large scales ends up when a final single vortex-like structure is formed,
as seen in the last snapshot taken at $t \simeq 34$. Simultaneously,
the disruptive influence of the magnetic field
causes small-scale MHD turbulence.

As in the two previous magnetic regimes there is a global increase
of the transverse kinetic energy following from
the large-scale coalescence events.
The disruptive effect can not change this dominant trend,
but partial magnetic reconnections allow the release of a non
negligible part of the energy built up by the instability at different
stages of the evolution.
From the transverse kinetic energy curve displayed in Fig.~\ref{tergnvor},
we can conclude that the scaling factor between the initial
small vortices and the final single one is now only of order $6$, while there
was a tenfold increase for the
very weak field as well as in the strong field case.
A similar time evolution
is obtained for a transonic layer with $M_A = 15$, and also for their supersonic
counterparts.

\subsection{The pairing/merging process}

As a white noise is used to initiate the instability, we need to explore
the sensitivity of the results to the initial conditions. To that end,
we performed three realisations ($A$, $B$, and $C$) of the same simulation.
This was done for the transonic layer with $M_A = 7$, in the disruptive regime
where the influence of small-scale reconnections is maximized.
The comparison of the time evolution of
the transverse kinetic energy is displayed in Fig.~\ref{abc}.
After a virtually identical linear phase (up to $t\approx 7$), all realizations
show a global increase for a extended period of time.
Hence, the overall scenario drawn in the previous subsection remains valid.
Differences in the time localization of the peaks are in agreement with
the different times for merging events.
The maximum amount of transverse kinetic
energy built up by the process differs from run to run.
The same is true for the transverse magnetic energy evolutions.
More precisely, the overall increase is fivefold, sixfold, and
ninefold for runs $A$, $B$, and $C$, respectively.
This means that the amount of energy release is
not the same and is very sensitive to the initial seed perturbation.
This is due to transient reconnection events, that are able to either partially
or almost completely release
built-up magnetic energy depending on the run considered.
This is also observed in the differences in density structure
(not shown) at intermediate times. Specifically, run $A$
has a more turbulent aspect than run $C$.

In order to understand in more detail
the pairing/merging process occurring between adjacent vortices, we
now turn to a more deterministic study of the interaction between
two identical vortices only. This is done using VAC,
setting $L_x = L_y = 2$, and taking a resolution of $400\times800$
grid cells. The case investigated is a
transonic shear layer with a magnetic field
strength corresponding to $M_A = 10$. We follow the time evolution
of the system, now perturbed by the deterministic disturbance form given
by Eq.~(\ref{deter}). However,
two wavelengths with the same amplitude are now excited, $\lambda_1 = L_x$
and $\lambda_2 = L_x / 2$. The latter and the former wavelength correspond
roughly to the linearly fastest growing mode and to its first subharmonic,
respectively.
In pure hydrodynamics, it has been shown
that two relative phase angles $\Phi$ between
the subharmonic and its
fundamental modes are of particular importance.~\cite{pat76}
These special angles are even and odd multiples of $\pi/2$. Thus, $2$
simulations with $\Phi = 0$, and $\pi /2$ are performed. An additional run,
in which only the fundamental mode is perturbed, is done in
order to serve as a reference case.

The results for the time evolution of the transverse magnetic energy component
are plotted in Fig.~\ref{fig15}
for the three cases. For $\Phi = 0$, the pairing/merging
event starts to occur before the saturation phase of the individual vortices.
This saturation for a single vortex is
typically observed at $t \simeq 4.5$.
Indeed, on the first snaphot of Fig.~\ref{fig16},
one can see that the two vortices are slightly displaced up and
down before being elongated. The pairing continues through a rotation one
around each other of the two vortices. Meanwhile, the vortices become elongated
and magnetic reconnection starts to occur in each vortex separately at
about $t\simeq 5$ (second snapshot). The pairing is being completed
and coincides with the second increase in the transverse energy evolution.
The disruption of the resulting bigger vortex after times $t\approx 15$
further releases the energy built up by the vortical motion.
Figure~\ref{fig15} indicates that the maximal transverse magnetic energy
reached is approximately twice its value obtained in the absence of
subharmonic growth (reference case in which pairing/merging
is absent).
This is also true for the characteristics of the final relaxed
state at the end of the simulation for $\Phi = 0$.

For $\Phi = \pi/2$, the difference with the
reference case without subharmonic excitation,
as seen in the first snapshot of Fig.~\ref{fig17}, is
to alternately strengthen and weaken the two vortices, without upward and
downward displacement as in the $\Phi=0$ case.
The pronounced increase of the energy is absent, and
an evolution similar to the reference case is now observed.
This agrees with the
absence of a pairing/merging event before or during the saturation of the two
vortices. Later, as the two vortices are quasi-independently disrupted
by magnetic reconnection, the weaker vortex tends
to be shredded by the stronger one. This explains
the somewhat slower decrease of the energy observed
at $t \simeq 16$, when comparing
to the reference case. The final state is rather similar to the reference
case but the relaxed shear layer is enlarged by a few percent.

These deterministic
runs clearly demonstrate the central role played by the growth
of subharmonics
of the fundamental mode on the pairing/merging process. Moreover,
the relative phase angle between the fundamental and the subharmonic mode
appears to be an important control parameter. In a simulation with many initial
vortices, we can thus easily understand how pairing events between adjacent
vortices can be more or less favored depending on the
different angle values. For white noise excitations, these are randomly
distributed.
In particular for the disruptive regime, this explains the extreme
sensitivity to the initial conditions. As an illustrative example,
Fig.~\ref{coalm1ma7long} shows snapshots of a transonic, $M_A=7$ shear layer
that differ from the simulation shown in Fig.~\ref{coalm1ma7} only in the
size of the simulated domain as well as in the initial random perturbation.
As we doubled the domain size and resolution, up to $22$ vortices are now formed
at early times. Qualitatively though, their trend to large-scale structure
formation, partially countered by small-scale reconnection, is similar.

In 2D hydrodynamics as well as in a 2D transverse MHD
configuration, the coalescence has been shown to be
a self-organization process with a selective decay of enstrophy,
allowing the relaxation to a nearly minimum enstrophy state.~\cite{miu99}
Attempts to explore the role of the enstrophy in our MHD parallel
configurations were unsuccesful. More precisely, in the very
weak field regime (nearly hydrodynamics), it was not surprising to
find a continuous decrease of the enstrophy with time.
However, a non monotonous time evolution of the enstrophy
was obtained for the
disruptive regime, making thus the importance of enstrophy
much less evident.

\subsection{Tearing-type reconnection in very weak field regime}

Finally, we turn to an interesting result obtained for the
transitional Mach number $M_A = 30$. This value separates the
disruptive from the very weak field regime identified for
a single vortex. Multiple vortex studies for much weaker field cases showed
essentially hydrodynamic behaviour with only the large-scale trend obviously
appearing in the density evolution (see Fig.~\ref{coalm1ma100}). Well in the
disruptive regime, this trend is somewhat opposed by the possibility to
disrupt individual vortices by reconnection (see Fig.~\ref{coalm1ma7}).

Here, we investigate the pairing effect in $M_A=30$ shear layers. Both
randomly initiated as well as deterministically excited runs demonstrated
the formation of magnetic islands at an intermediate stage in their evolution.
This is illustrated in Fig.~\ref{islands}, where small islands are
clearly seen to grow at the periphery of the vortices in the density snapshot at $t=12$.
In this run, we maximized the effect by choosing a deterministic excitation
with zero phase differences between the various modes. We also use AMRVAC
setting $L_x = 8$ and $L_y = 4$.
The features are very similar to results obtained in
the non linear development of a single vortex
in the 2D KH unstable shear layer in the
presence of initially antiparallel magnetic fields
(see Fig.~9 of Keppens {\it et al.}~\cite{kep99}).
This island formation appears as
a result of a tearing-type instability induced by the vortical motions.
For the small initial magnetic field strength case of Fig.~\ref{islands},
the vortical motions associated to the initial small eddies are able
to roll-up the magnetic field lines in more
than one turn without disruptive effect. Thus, the subsequent pairing/merging
between two vortices lead to push antiparallel field lines together, forming
thin current sheets. These eventually become unstable and trigger
magnetic islands at the periphery of the resulting bigger vortices.
The result shown in Fig.~\ref{islands} is in fact obtained for a fully
resistive MHD simulation, where the constant resistivity coefficient is
$\eta=3.33 \times 10^{-5}$. At the effective resolution of this AMR simulation,
this low value for the resistivity is just dominating the numerical dissipation
in the calculation (see the convergence study in Keppens
{\it et al.}~\cite{kep99}). Note that
the reconnection events cause a rapid transition to MHD turbulence, superposed
on the large scale vortex structures.

\section{Summary, outlook and astrophysical relevance}

We can summarize our findings as follows. We have numerically studied the 
development of the KH instability that occurs in a 2D parallel magnetized shear 
flow layer. We extend previous studies made for a single periodic vortex in 
subsonic/transonic layers,
to a configuration allowing the growth of many linearly dominant wavelengths
along the layer.
First, for a slightly supersonic layer having a sonic Mach number $M_s = 1.4$
and a subsonic convective Mach number $M_{cs} = 0.7$, we
confirm the existence of three dynamically different regimes according to 
the relative magnetic field strength.
Indeed, for a strong enough magnetic field with 
$2 \simlt M_A \simlt 4$,
the KH instability is halted by the magnetic field tension generated by the
vortical motion itself. In the opposite regime, for an Alfv\'en number exceeding
a critical value $M_c$, $M_A \ge M_c$, the development of the vortex
is hydrodynamics-like. The intermediate regime is the most interesting one from
the point of view of the dynamics. Indeed, the magnetic field structure generated by 
the vortical motion is able to saturate and disrupt the vortex due to magnetic 
reconnection. The high number of runs allow us to quantify the transitions 
between the different regimes. In particular, we obtained evidence that 
the transitional Alfv\'en Mach
number value $M_c$ separating the disruptive 
from the very weak field regimes is closer
to $M_c = 30$ than to $M_c = 20$, as previously determined. Moreover, the 
disruptive regime can be separated into two subregimes, according to $M_A$ 
greater or smaller than $8$. This corresponds to the rolling-up value of 
the interface at saturation,
that is $1/2$, $1$, and $2$ turns for $M_A = 5, 8$, and $30$, 
respectively. The comparison
of results obtained for $M_s = 1.4$ and $M_s = 1$ show many similarities, 
except that the maximum density contrast is higher by approximately $50$ 
percent for the supersonic layer.
An additional feature is the formation of
eddy shocklets of fast magnetosonic type. These are transient for
$M_A = 15$ and persistent for $M_A = 100$. 
For truly supersonic flow ($M_s = 3$, where also $M_{cs}>1$),
the dominant instability changes character, having both a low linear growth
rate and a low saturation level. The associated unstable traveling modes for
highly supersonic layers are much less disruptive for the flow.

Second, the use of adaptive mesh refinement in our code allowed
us to explore large spatial domains containing transonic and slightly supersonic
shear layers. A white noise perturbation is added to the background flow in order
to let the most unstable modes grow in a natural way. Typically, up to $22$
associated wavelengths are initially observed to grow in accord with
the linear theory.
In the very weak field regime,
a continuous sequence of pairing/merging events between 
vortices is obtained, that
ends only when a single big long-lived vortex is formed.
This is very similar to the behaviour known
in pure 2D hydrodynamics.~\cite{les88} This trend towards large-scale
coalescence is also observed for the disruptive regime, but it is accompanied
by magnetic reconnection events that are able to partially disrupt the 
vortices at different stages of the evolution.
We have also demonstrated the sensitivity of the results to the seed perturbation,
that is stronger than for a purely hydrodynamic regime.
This is a consequence of the central role played by
the relative phase differences between the subharmonic modes,
that are responsible for the large-scale coalescence. For a particular magnetic field
strength of the very weak field regime ($M_A = 30$), tearing-type reconnection events
are identified to occur during the cascade towards large scales.
When the magnetic field is so strong (strong field regime) as to
prevent the formation of vortices, the large-scale coalescence is once more obtained
after a relaxation phase. Therefore, we can conclude that the coalescence towards
large scales is a strong mechanism, that can not be stopped
by disruptive effects. Magnetic reconnection can only partially release
the energy built up by the whole instability mechanism.

It will be of interest to study whether the trend to form a large-scale structure
eventually ceases when a certain longitudinal lengthscale is reached. A critical
factor will be the dependence of the cross-stream scaling factor relating the
characteristics of the bigger vortices formed through mergings with the
original smaller ones. So far, in none of our simulations covering up to
22 wavelengths, has the merging ceased before reaching the full computational
domain size.

Finally, the issue of large-scale, magnetized flow coherence and
survival is of fundamental importance for astrophysical jets.
Indeed, very high resolution hydrodynamic simulations are unable to
reproduce the remarkable stability deduced from observations.~\cite{bod95,bod98}
Attempts to stabilize such highly supersonic jets, invoke
jet densities much higher than that of the surrounding medium and/or
favorable radiative effects (critically dependent on the choice of the cooling
function).~\cite{dow98,mic00,sto97}
However, as the presence of non negligible magnetic fields is necessary to ensure
the collimation of jets, the solution to this problem could also be magnetic.
A recent numerical study has shown that a cylindrical jet will likely be
subject to both KH and current-driven modes. These latter instabilities are
of magnetic origin, and result from the presence of helical magnetic fields.
It has been obtained that the nonlinear interaction between
simultaneously growing KH and current-driven modes can in fact 
aid jet survival.~\cite{bat02} 
Without invoking magnetic instabilities, MHD simulations of the long-term 
evolution
of a whole jet configuration have shown that jets embedded in a helical magnetic field
seem to be significantly more stable than similar flows in a
purely axial field.~\cite{ros99}
The enhanced linear stability due to the azimuthal field component 
that is predicted by theory, is probably not sufficient to explain this effect.~\cite{app92}
This indicates the necessity to investigate in detail all aspects occurring
in the nonlinear regime.
The large-scale coalescence observed in the present work could indeed play
an important role, as this mechanism continuously transfers 
the free energy towards the large scales
without fully releasing it in a disruptive way. 
In three dimensional hydrodynamics,
this is not possible as 3D instabilities are able to break up the jet in a
turbulent transition, in which all the important physical quantities cascade toward high
wavenumbers until they dissipate.~\cite{les97} However, in MHD, a strong 
inverse
cascade toward small wavenumbers is allowed both in 2D and 3D. Indeed, self-organization
processes, which lead to the formation of large-scale coherent structures, follow from
the inverse cascade of the mean square magnetic potential and magnetic helicity in 2D
and 3D, respectively.~\cite{bis93} A representative 3D example of such an effect is the
subsequent formation of a large-scale magnetic field in association with a 
dynamo mechanism.~\cite{bra01}
Hence, the large-scale coalescence effect should be investigated in more
detail for 3D jet-like magnetized configurations. 

\begin{acknowledgements}
This work was supported in part by the European Community's
Human Potential Programme under contract HPRN-CT-2000-00153, PLATON.
RK performed this work in part supported 
by the European Communities under the contract of
Association between Euratom/FOM, carried out within the framework of the
European Fusion Programme. Views and opinions expressed herein do not 
necessarily reflect those of the European Commision. RK acknowledges the 
Computational Science programme `Rapid Changes in Complex Flow' funded by
NWO-E, coordinated by J.P. Goedbloed. HB thanks J.P. Goedbloed for
his hospitality during his visit to the FOM institute. NCF is acknowledged
for computing facilities. 
We thank an anonymous referee for useful suggestions.
\end{acknowledgements}

\newpage
{\bf Figure Captions:}

\vspace*{0.5cm}
{\bf Figure 1:}
Time evolution of the transverse kinetic $e_{k,y}$ (upper panels), transverse
magnetic $e_{m,y}$ (middle panels), and the longitudinal kinetic $E_{k,x}$ energies (bottom panels)
for a strong field case $M_A = 3.33$ (dash-dotted line), 
a disruptive case $M_A = 10$ (solid line), and a very weak field case
$M_A = 100$ (dashed line). Shown are transonic $M_s = 1$ (left panels)
and supersonic $M_s = 1.4$ (right panels) cases.

\vspace*{0.5cm}
{\bf Figure 2:}
Saturation level versus the Alfv\'en Mach number $M_A$ for the
transonic $M_s = 1$ (circles), and supersonic $M_s = 1.4$ (squares) cases.
We use the first maximum of the transverse
kinetic $e_{k,y}$ and magnetic $e_{m,y}$
energies in (a) and (b), respectively. The transitions between the three
magnetic regimes are indicated.

\vspace*{0.5cm}
{\bf Figure 3:}
Colour images of the density distribution at saturation, for
the transonic $M_s = 1$ case (top panels) versus the supersonic $M_s=1.4$
case (bottom panels). Shown are density maps for
Alfv\'en Mach numbers $M_A=15$ and $100$. Only a part with $y$ in the range
$[-0.35:0.35]$ of the full grid is shown.

\vspace*{0.5cm}
{\bf Figure 4:}
Left: Cross-sectional variation of various quantities through the upper shock 
taken at $t=4.26$ (as in Fig.~\ref{density}) 
and at $y=0.3$ for the $M_A=15$ case.
Right: Similarly for the $M_A=100$ case at $t=4.26$ (slightly later
than in Fig.~\ref{density} where $t=4$) and at $y=0.2$.

\vspace*{0.5cm}
{\bf Figure 5:}
Time evolution of the longitudinal $E_{k,x}$ (plain line) and transverse 
$e_{k,y}$ (dotted line) kinetic energies, for transonic $M_s=1$
and three Alfv\'en $M_A$ Mach numbers. The values are normalized by the initial
value of $E_{k,x}$ at $t=0$. The transverse energy is additionally multiplied by
a scale factor that is equal to $30$ for $M_A = 3.33, 7$, and to $7$ for $M_A = 100$.

\vspace*{0.5cm}
{\bf Figure 6:}
Grey-scale images of the density distribution of a very weak field
transonic layer with $M_A = 100$. The contour levels are normalized
using a linear scale with density values ranging from $0.42$ to $1.13$.
Only a part with $y$ in the range $[-2:2]$ of the full grid is shown.
Times are indicated at left.

\vspace*{0.5cm}
{\bf Figure 7:} Contour levels of the density for 2 snapshots in the
evolution of a supersonic layer $M_s = 1.4$
with $M_A = 100$. Also indicated is the location of the finest level grids
in the grid-adaptive simulations: note how the shock fronts are fully
captured at the highest resolution.

\vspace*{0.5cm}
{\bf Figure 8:}
Same as Fig.~\ref{coalm1ma100}
but for a transonic layer $M_s =1$ with a strong field with $M_A = 3.33$.
The contour levels are normalized using a linear scale with density values
ranging from $0.67$ to $1.12$.

\vspace*{0.5cm}
{\bf Figure 9:}
Same as Figure~\ref{coalm1ma100} but for a transonic layer $M_s =1$ 
with $M_A = 7$ (disruptive).
The contour levels are normalized using a linear scale with density values
ranging from $0.52$ to $1.21$.

\vspace*{0.5cm}
{\bf Figure 10:}
Time evolution of the transverse kinetic energy $e_{k,y}$
for a transonic shear flow layer $M_s = 1$, with $M_A = 7$. We display
three realisations A (dash-dotted line), B (plain line), and C
(dashed line) of the same physical case, only differing in the initial random
perturbation.

\vspace*{0.5cm}
{\bf Figure 11:}
Time evolution of the transverse magnetic energy $e_{m,y}$ of two
identical interacting vortices
for a transonic $M_s = 1$ layer. A value $M_A = 10$
is taken. The initial perturbation has a relative phase difference $\Phi = 0$
(plain line), and $\Phi = \pi/2$ (dashed line). A simulation with only the
fundamental perturbed is also reported (dash-dotted line).

\vspace*{0.5cm}
{\bf Figure 12:}
Grey-scale images of the density distribution of a transonic layer
corresponding to Fig.~\ref{fig15} with a relative perturbation angle $\Phi = 0$.
The contour levels are normalized
using a linear scale with density values ranging from $0.51$ to $1.10$.

\vspace*{0.5cm}
{\bf Figure 13:}
Grey-scale images of the density distribution of a transonic layer
corresponding to Figure~\ref{fig15} with a relative
perturbation angle $\Phi = \pi/2$.
The contour levels are normalized
using a linear scale with density values ranging from $0.54$ to $1.09$.

\vspace*{0.5cm}
{\bf Figure 14:}
Different realization of
a transonic layer $M_s =1$ with $M_A = 7$,
as in Figure~\ref{coalm1ma7}, but for a double domain size (and resolution) including up to 22
vortices at early times. We set $L_x = 20$ and $L_y = 8$. Only a part with $y$ in the range
$[-2:2]$ of the full grid is shown.

\vspace*{0.5cm}
{\bf Figure 15:}
Density evolution in a transonic layer $M_s =1$ with $M_A = 30$.
The contour levels are normalized using a linear scale with density values
ranging from $0.61$ to $1.12$. Note the formation of small magnetic 
islands, nicely
visible in the density structure at $t=12$ (fourth snapshot)
at the periphery of the vortices.

\end{document}